\RequirePackage[british]{babel}
\documentclass[reqno,a4paper,10pt]{amsart}
\usepackage[utf8x]{inputenc}
\usepackage[hmargin=3.3cm]{geometry}
\usepackage{amsmath,amssymb,amstext,amsthm,amscd,mathrsfs,eucal}
\usepackage{graphicx,color}
\usepackage{array}
\usepackage{hyperref}
\hypersetup{%
  pdftitle   = {Half-BPS quotients in M-theory: ADE with a twist},
  pdfkeywords = {AdS/CFT, M2, M-theory, Killing spinors, 7-sphere quotients},
  pdfauthor  = {Paul de Medeiros, José Figueroa-O'Farrill, Sunil Gadhia, Elena Méndez-Escobar},
  pdfcreator = {\LaTeX\ with package \flqq hyperref\frqq}
}
\PrerenderUnicode{éÉ}
\let\into\hookrightarrow
\newcommand{\half}{\tfrac12}

\newcommand{\Cl}{\mathrm{C}\ell}

\newcommand{\rv}{\mathsf{\Lambda}^1}
\newcommand{\rs}{{\mathsf{\Delta}_+}}
\newcommand{\rc}{{\mathsf{\Delta}_-}}
\newcommand{\rsc}{{\mathsf{\Delta}_\pm}}

\newcommand{\sA}{\mathsf{A}}
\newcommand{\sD}{\mathsf{D}}
\newcommand{\sE}{\mathsf{E}}

\newcommand{\I}{\mathfrak{i}}
\newcommand{\J}{\mathfrak{j}}
\newcommand{\K}{\mathfrak{k}}

\newcommand{\fso}{\mathfrak{so}}
\newcommand{\fosp}{\mathfrak{osp}}
\newcommand{\fsp}{\mathfrak{sp}}
\newcommand{\fsu}{\mathfrak{su}}
\newcommand{\fu}{\mathfrak{u}}

\newcommand{\SO}{\mathrm{SO}}
\newcommand{\Spin}{\mathrm{Spin}}
\newcommand{\Sp}{\mathrm{Sp}}
\newcommand{\SU}{\mathrm{SU}}
\newcommand{\U}{\mathrm{U}}

\newcommand{\PSL}{\mathrm{PSL}}

\newcommand{\SL}{\mathrm{SL}}
\newcommand{\GL}{\mathrm{GL}}

\newcommand{\FF}{\mathbb{F}}
\newcommand{\PP}{\mathbb{P}}
\newcommand{\RR}{\mathbb{R}}
\newcommand{\HH}{\mathbb{H}}
\newcommand{\CP}{\mathbb{CP}}
\newcommand{\RP}{\mathbb{RP}}
\newcommand{\CC}{\mathbb{C}}
\newcommand{\ZZ}{\mathbb{Z}}

\DeclareMathOperator{\AdS}{AdS}
\DeclareMathOperator{\Aut}{Aut}

\DeclareMathOperator{\Hom}{Hom}

\DeclareMathOperator{\Tr}{Tr}

\theoremstyle{plain}

\theoremstyle{definition}

\newcommand{\MUNCH}[1]{\relax}

\allowdisplaybreaks
\begin{document}
\title[Half-BPS quotients in M-theory]{Half-BPS quotients in M-theory: ADE with a twist}
\author[de Medeiros]{Paul de Medeiros}
\author[Figueroa-O'Farrill]{José Figueroa-O'Farrill}
\author[Gadhia]{Sunil Gadhia}
\author[Méndez-Escobar]{Elena Méndez-Escobar}
\address{School of Mathematics and Maxwell Institute for Mathematical Sciences, University of Edinburgh, James Clerk Maxwell Building, King's Buildings,
  Edinburgh EH9 3JZ, UK}
\address[JMF also]{Institute for the Physics and Mathematics of the Universe, University of Tokyo, Kashiwa, Chiba 277-8586, Japan}
\email{\{P.deMedeiros,J.M.Figueroa,E.Mendez\}@ed.ac.uk, sunilgadhia@googlemail.com}
\begin{abstract}
  We classify Freund--Rubin backgrounds of eleven-dimensional supergravity of the form $\AdS_4 \times X^7$ which are at least half BPS --- equivalently, smooth quotients of the round 7-sphere by finite subgroups of $\SO(8)$ which admit an ($N>3$)-dimensional subspace of Killing spinors.  The classification is given in terms of pairs consisting of an ADE subgroup of $\SU(2)$ and an automorphism defining its embedding in $\SO(8)$.  In particular we find novel half-BPS quotients associated with the subgroups of type $\sD_{n\geq 6}$, $\sE_7$ and $\sE_8$ and their outer automorphisms.
\end{abstract}
\maketitle

\section{Introduction and contextualisation}
\label{sec:introduction}

Recent progress in the AdS/CFT correspondence for M2-branes \cite{Malda} makes desirable a precise dictionary between the near-horizon geometries of supersymmetric M2-branes and their dual three-dimensional superconformal field theories.  The emerging picture deriving from the pioneering work of Bagger and Lambert \cite{BL1,BL2}, Gustavsson \cite{GustavssonAlgM2} and Aharony, Bergman, Jafferis and Maldacena \cite{MaldacenaBL} is that the dual theories are superconformal Chern--Simons theories coupled to matter.  Superconformal field theories in three-dimensions are invariant under the orthosymplectic Lie superalgebra $\fosp(N|4)$ \cite{Nahm}, where $1\leq N \leq 8$.  As reviewed for example in \cite{SCCS3Algs}, which contains a more comprehensive bibliography, superconformal Chern--Simons theories can be labelled by a pair consisting of a metric Lie algebra and a unitary representation, with a precise condition on the type of representations allowed/required for each value of $N$.  For $N>3$ the representation theoretic data determining the superconformal Chern--Simons theory is very constrained and a classification is possible.  This classification is very concrete for $N>4$ and not yet completely explicit in the case of $N=4$; although there is a clear algorithm how to construct all such theories.  It is therefore for the $N>3$ theories that we expect the dictionary to be easiest to establish.  As a preliminary step in this programme, one needs a classification of the possible near-horizon geometries at least for $N>3$ and the purpose of this note is precisely to complete this classification.

As explained, for example, in \cite{AFHS,MorrisonPlesser}, the near-horizon geometry of a supersymmetric M2-brane configuration admitting a superconformal field theory is of the form $\AdS_4 \times X^7$, where $(X,g)$ is a riemannian 7-dimensional manifold admitting real Killing spinors.  This is the near-horizon geometry of an M2-brane background where the transverse space to the membrane is the metric cone $C(X)$ of $X$.  As a riemannian manifold, $C(X) = \RR^+ \times X$ with metric $g_C = dr^2 + r^2 g$, where $r > 0$ is the coordinate on $\RR^+$.  Bär's cone construction \cite{Baer} says that $(X,g)$ admits real Killing spinors if and only if $(C(X),g_C)$ admits parallel spinors.  If $X$ is complete, then a theorem of Gallot's \cite{Gallot} says that $C(X)$ is either flat or irreducible.  The irreducible holonomy representations in eight dimensions follow from Berger's list and M.~Wang \cite{Wang} determined which of them admit parallel spinors.  The well-known answer is that an irreducible, simply-connected, riemannian eight-dimensional manifold admitting parallel spinors must have precisely one of the following holonomy groups: $\Spin(7)$, $\SU(4)$ or $\Sp(2)$, the last two corresponding to Calabi--Yau 4-folds and hyperkähler 8-manifolds, respectively.  Letting $N$ denote the dimension of the space of parallel spinors (of, say, positive chirality): we find that $\Spin(7)$-holonomy manifolds have $N=1$, whereas Calabi--Yau 4-folds have $N=2$ and hyperkähler manifolds have $N=3$.  If the manifolds are not simply connected, then of course $N$ may be lower.  It follows that for $N>3$ the holonomy has to be reducible, but then Gallot's theorem says that the cone must be flat, hence a quotient of $\RR^8$.  As the notation suggest, the integer $N$ in this paragraph is the same as in the previous paragraph.  It can also be interpreted in terms of the fraction of supersymmetry preserved by the near-horizon geometry of the M2-brane solution, which is $\frac{N}{8}$.  The geometric construction of the $\fosp(N|4)$ algebra starting from the supergravity background was given in \cite{JMFKilling}.

In terms of the 7-dimensional manifold $X$, the reduced holonomy of the cone implies the existence of parallel differential forms which, when contracted with the Euler vector $r\frac{\partial}{\partial r}$, give rise to geometric objects on $X$ satisfying a number of identities defining certain well-known classes of geometries: manifolds of weak $G_2$ holonomy, Sasaki-Einstein and 3-Sasaki 7-manifolds, respectively, for the irreducible cones.  If the cone is flat, then $X$ is locally isometric to the round 7-sphere $S^7$. \emph{We will assume that $X$ is smooth.}  Therefore the manifolds $X$ for which $N>3$ are quotients $S^7/\Gamma$ where $\Gamma< \SO(8)$ is a finite group acting freely on $S^7$ in such a way that the quotient is spin and admits an ($N>3$)-dimensional space of Killing spinors.

There is some previous literature on this problem, which is reviewed in \cite[§3.1]{MorrisonPlesser} (and references therein).  The case of $S^7/\Gamma$ a lens space has also been studied in \cite{Franc} and \cite{BaerSpheres} and more recently also in \cite{FigGadLens} emphasising the rôle of the spin structure.  This corresponds to $\Gamma$ a finite cyclic group and all examples of $N>5$ are of this form.  In \cite[§3.1]{MorrisonPlesser} there is a brief discussion of $N=5$ examples and also some $N=4$ lens spaces.  In this note we complete the case of $N=4$ and hence the $N>3$ classification.

The organisation of this note is very simple.  After a brief introduction to the concrete mathematical problem in Section~\ref{sec:spherical-quotients}, we discuss the quotients $S^7/\Gamma$ with $N>3$ in turn, starting with $N=8$ and ending with $N=4$, this latter case taking up the bulk of the note.  We end by summarising the results in Table \ref{tab:quotients}.  In particular, there are examples consisting of twisted (in a sense to be made precise below) embeddings of the binary dihedral, binary octahedral and binary icosahedral groups which appear to be new.

\section{Sphere quotients with Killing spinors}
\label{sec:spherical-quotients}

We now describe the classification of smooth quotients $S^7/\Gamma$ of the unit sphere in $\RR^8$ by finite subgroups $\Gamma < \SO(8)$, for which the space of real Killing spinors has dimension $N>3$.  Bär's cone construction reduces the classification of these quotients to the classification of finite subgroups $\Gamma < \Spin(8)$ acting freely on the unit sphere in the vector representation $\rv$ and leaving invariant (pointwise) an $N$-dimensional subspace in either of the half-spinor representations $\rsc$.  The subgroups of $\SO(8)$ which act freely on the unit sphere have been classified by Wolf \cite{Wolf} and their lifts to $\Spin(8)$ have been classified in \cite{SunilThesis}.  Every lift is a choice of spin structure on the quotient and this choice will manifest itself very concretely as signs we can attach to the generators of $\Gamma$ consistently with the relations.  This is to be expected, since spin structures on $S^7/\Gamma$ are classified by
\begin{equation}
  \label{eq:spinstructs}
  H^1(S^7/\Gamma,\ZZ_2) \cong \Hom(\pi_1(S^7/\Gamma),\ZZ_2) \cong \Hom(\Gamma,\ZZ_2) \cong \Hom(\Gamma^{\text{ab}}, \ZZ_2)~,
\end{equation}
where the abelianisation $\Gamma^{\text{ab}}$ of $\Gamma$ is the quotient of $\Gamma$ by its commutator subgroup $[\Gamma,\Gamma]$ consisting of elements of the form $aba^{-1}b^{-1}$ for $a,b\in\Gamma$.

Let $\Gamma < \Spin(8)$ be a subgroup not containing $-1$.  This condition guarantees that it maps isomorphically under $\Spin(8) \to \SO(8)$ to some subgroup (also denoted) $\Gamma<\SO(8)$.  Assume that it acts freely on $S^7 \subset \rv$.  The quotient $S^7/\Gamma$, which is thus smooth and spin, is said to be of type $(N_+,N_-)$, where  $N_\pm = \dim {\rsc}^\Gamma$.

There are two approaches one can take to attack this problem.  As in \cite{SunilThesis}, one can start from Wolf's list of smooth $S^7$ quotients, analyse the possible lifts to $\Spin(8)$ and compute the dimension of the invariant subspace in the spinor representation.  Alternatively, one can start from finite subgroups of $\Spin(8)$, not containing $-1$ and which leave invariant a subspace of spinors of the requisite dimension and then check whether the action on $S^7$ is free.  This  latter approach works well for $N>3$, but perhaps not so well for smaller values of $N$.

Indeed, suppose that $\Gamma < \Spin(8)$, not containing $-1$, leaves invariant an $N$-dimensional subspace $V\subset\rs$ and let $\rs = V \oplus V^\perp$. Let $\SO(V^\perp)$ denote the subgroup of $\SO(\rs)$ which acts trivially on $V$.  Then $\Gamma$ belongs to the preimage of $\SO(V^\perp)$ in $\Spin(8)$, which is isomorphic to $\Spin(8-N)$.  The group $\Spin(8)$ acts transitively on the grassmannian of $N$-planes in $\rs$, whence any two $\SO(V^\perp)$ subgroups will lift to $\Spin(8-N)$ subgroups of $\Spin(8)$ which are conjugate, and quotients by conjugate subgroups are isometric.  It is therefore enough to fix one $\Spin(8-N)$ subgroup of $\Spin(8)$ once and for all.  In summary, the problem of classifying quotients of type $(N,0)$ is tantamount to classifying finite subgroups of a fixed $\Spin(8-N)$ subgroup of $\Spin(8)$ whose action on the unit sphere of the vector representation $\rv$ is free.  The reason this method works well for $N>3$ is that we have good control over the finite subgroups of $\Spin(2) \cong \U(1)$, $\Spin(3) \cong \Sp(1)$ and hence also of $\Spin(4) \cong \Sp(1) \times \Sp(1)$.

We now proceed to list the smooth quotients with $N>3$.  One first fixes an $N$-dimensional subspace $V \subset \rs$ and identifies the Lie subalgebra of $\fso(8)$ which leaves $V$ pointwise invariant.  This calculation is easily performed in the Clifford algebra $\Cl(8)$ relative to an explicit matrix realisation, for which it is useful to use Mathematica.  Exponentiating in $\Cl(8)$ determines a $\Spin(8-N)$ subgroup whose action on the unit sphere $S^7 \subset \rv$ in the vector representation can then be easily investigated.  Our Clifford algebra conventions are as follows: the generators of $\Cl(8)$ are $\gamma_i$, for $i=1,2,\dots,8$ and satisfy $\gamma_i^2=-1$ in our conventions.  We use the shorthand $\gamma_{ij\cdots k} =\gamma_i \gamma_j \cdots \gamma_k$ for $i,j,\ldots,k$ distinct. We will employ an explicit realisation based on the octonions in which $\gamma_i$ are real, skewsymmetric $16\times 16$ matrices whose entries take only the values $0,\pm1$.

It will be convenient to introduce the following elements in $\Cl(8)$:
\begin{equation}
  \label{eq:IJK}
  \begin{aligned}[m]
    \I &= \half( \gamma_{12} + \gamma_{34})\\
    \J &= \half( \gamma_{13} - \gamma_{24})\\
    \K &= \half( \gamma_{14} + \gamma_{23})\\
  \end{aligned}
  \qquad\qquad
  \begin{aligned}[m]
    \I' &= \half( \gamma_{56} + \gamma_{78})\\
    \J' &= \half( \gamma_{57} - \gamma_{68})\\
    \K' &= \half( \gamma_{58} + \gamma_{67})\\
  \end{aligned}
\end{equation}
Each of $\I$, $\J$, $\K$ squares to $-P_-$ and similarly with primes, where $P_- = \half(1-\gamma_{1234})$ and $P'_- = \half(1-\gamma_{5678})$.  These elements span
an $\fsp(1) \oplus \fsp(1)$ Lie subalgebra of the $\fso(8)$ Lie algebra spanned by the $\gamma_{ij}$, which the exponential maps surjectively onto an $\Sp(1) \times \Sp(1)$ Lie subgroup of $\SO(8)$ where all the finite groups we shall consider lie.

\section{$N=8$}

The (unquotiented) 7-sphere possesses a unique spin structure admitting the maximal number of Killing spinors of either sign of Killing's constant, so it has type $(8,8)$, since $\dim\rsc = 8$.  The real projective space $\RP^7$ is the quotient of the sphere by a $\ZZ_2$ subgroup of $\SO(8)$ which sends $x \in \RR^8$ to $-x$.  This subgroup admits two inequivalent lifts to $\Spin(8)$: given by $\pm \gamma_9$, where $\gamma_9$ is the chirality operator in $\Cl(8)$.  This means that $\RP^7$ has two inequivalent spin structures: one of type $(8,0)$ and the other of type $(0,8)$.  Any other quotient will have type $(N,0)$ or $(0,N)$ for some $N<8$.  Up to change of orientation, we may consider only those quotients of type $(N,0)$.

\section{$N=7$}

There is no quotient with $N=7$.  This is proved by an argument employed in \cite{FigGadPreons} in a related context.  Suppose $\Gamma < \Spin(8)$ leaves invariant (pointwise) a 7-dimensional subspace of $\rs$, then so will some element $1 \neq g \in \Gamma$.  Now $g$ generates a cyclic subgroup of order $>1$, lying inside some circle group inside a maximal torus.  The only such freely acting circle groups have as orbits the fibres of the canonical fibration $S^7 \to \CP^3$.  That group preserves a complex structure on $\rs$, whence the invariant subspace is a complex subspace and hence always of even dimension.

\section{$N=6$}

Groups $\Gamma$ leaving invariant a six-dimensional subspace of $\rs$ are contained in a circle subgroup of $\Spin(8)$ whose Lie algebra is conjugate to the $\fu(1)$ generated by $\I + \I'$, where we view the spin group and its Lie algebra as sitting inside the Clifford algebra $\Cl(8)$.  The commutant of $\fu(1)$ is $\fu(1) \oplus \fsu(4)$, where $\fsu(4) \cong \fso(6)$ is the Lie algebra of the $\Spin(6)$ subgroup of $\Spin(8)$ associated to the six-dimensional vector space of invariant spinors.  The finite subgroups of a circle group are cyclic and it is easy to check that they act freely on $S^7$: they rotate all elementary 2-planes by the same angle.  Alternatively, thinking of $\rv$ as $\CC^4$, they act by multiplying each component by the same phase, whence only the origin is fixed, but the origin does not lie on the unit sphere.

\section{$N=5$}

Groups $\Gamma$ leaving invariant (pointwise) a five-dimensional subspace of $\rs$ are contained in a $\Spin(3)$ subgroup of $\Spin(8)$ whose action on the vector representation is obtained as follows.  It will prove convenient to employ the isomorphism $\Spin(3) \cong \Sp(1)$ with the unit quaternions.  Consider the action of $\Sp(1)$ on $\HH$ by left quaternion multiplication: if $u \in \Sp(1)$ and $q \in \HH$, then $u \cdot q = uq$.  Thinking of $\HH$ as a four-dimensional real vector space with basis $1,i,j,k$, we get an embedding $\Sp(1) \into \SO(4)$ given explicitly by
\begin{equation}
\label{eq:sp1inso4}
\zeta_1 + \zeta_2 i + \zeta_3 j + \zeta_4 k \mapsto
  \begin{pmatrix}
    \phantom{-}\zeta_1 & -\zeta_2 & -\zeta_3 & -\zeta_4 \\
    \phantom{-}\zeta_2 & \phantom{-}\zeta_1 & -\zeta_4 & \phantom{-}\zeta_3 \\
    \phantom{-}\zeta_3 & \phantom{-}\zeta_4 & \phantom{-}\zeta_1 & -\zeta_2 \\
    \phantom{-}\zeta_4 & -\zeta_3 & \phantom{-}\zeta_2 & \phantom{-}\zeta_1
  \end{pmatrix}~,
\end{equation}
and we can them embed $\SO(4)$ diagonally into $\SO(4)\times \SO(4) \subset \SO(8)$.  In other words, thinking of $\RR^8$ as $\HH^2$, if $u \in \Sp(1)$ is a unit quaternion, then its action on $(q_1,q_2) \in \HH^2$ is simply via left multiplication: $u \cdot (q_1, q_2) = (u q_1, u q_2)$.  It is clear that this action is free on the sphere, as the only point $(q_1,q_2)$ which has a nontrivial stabilizer is the origin, which does not lie on the unit sphere.

The finite subgroups of $\Sp(1)$ are classified by the ADE Dynkin diagrams and tabulated in Table~\ref{tab:ADE}.  The cyclic groups corresponding to $\sA_{n\geq 2}$ actually leave invariant a six-dimensional subspace and are precisely the groups considered above in the section on $N=6$.  The cyclic group of order $2$ corresponding to $\sA_1$ actually acts trivially on $\rs$ and is the group whose quotient is the real projective space discussed in the section on $N=8$.

\begin{table}[h!]
  \caption{Finite subgroups of $\Sp(1)$}
  \centering
  \begin{tabular}{>{$}c<{$}|>{$}c<{$}|>{$}c<{$}|>{$}l<{$}}
    \text{Dynkin} & \text{Name} & \text{Order} & \text{Presentation}\\\hline
    \sA_n & \ZZ_{n+1} & n+1 & \left<t \middle| t^{n+1}=1\right>\\
    \sD_{n\geq 4} & 2D_{2(n-2)} & 4(n-2) & \left<s,t \middle| s^2=t^{n-2}=(st)^2\right>\\
    \sE_6 & 2T & 24 & \left<s,t \middle| s^3=t^3=(st)^2\right>\\
    \sE_7 & 2O & 48 & \left<s,t \middle| s^3=t^4=(st)^2\right>\\
    \sE_8 & 2I & 120 & \left<s,t \middle| s^3=t^5=(st)^2\right>
  \end{tabular}
  \label{tab:ADE}
\end{table}

\section{$N=4$}

Groups $\Gamma$ leaving invariant (pointwise) a four-dimensional subspace of $\rs$ are contained in a $\Spin(4) \cong \Sp(1) \times \Sp(1)$ subgroup of $\Spin(8)$ whose action on the vector representation is defined in a very similar way to the case of $N=5$, except that we have two copies of $\Sp(1)$, each embedding in $\SO(4)$ as in \eqref{eq:sp1inso4} and then embedding $\SO(4) \times \SO(4)$ into $\SO(8)$ in the obvious way.  In terms of quaternions, $(u_1,u_2) \in \Sp(1) \times \Sp(1)$ acts on $(q_1,q_2) \in \HH^2$ by left multiplication: namely, $(u_1,u_2) \cdot (q_1,q_2) = (u_1 q_1, u_2 q_2)$.  Now consider a finite subgroup $\Gamma < \Sp(1) \times \Sp(1)$.  It will consist of pairs $(u_1,u_2)$ of unit quaternions acting via simultaneous left multiplication on pairs of quaternions.  It is clear that for this action to be free on the sphere, it is necessary and sufficient that there be no elements in $\Gamma$ of the form $(1,u)$ or $(u,1)$, other than the identity $(1,1)$.  To see what this means, let's digress to review Goursat's theory of subgroups of the direct product of two groups \cite[§4.3]{MR1957212}.

Let $A,B$ be groups and let $C < A \times B$ be a subgroup.  Let $L = \pi_1(C) < A$ and $R = \pi_2(C) < B$ be the images of $C$ under the cartesian projections $\pi_1 : A\times B \to A$ and $\pi_2 : A \times B \to B$, respectively.  Alternatively, $L<A$ consists of those $a \in A$ such that there exists $b \in B$ with $(a,b) \in C$, and similarly $R < B$ consists of those $b \in B$ such that there exists $a \in A$ with $(a,b) \in C$.  Let $L_0<L$ consist of those $a \in L$ such that $(a,1) \in C$, and let $R_0<R$ consist of those $b \in R$ such that $(1,b) \in C$.  It is easy to see that $L_0\lhd L$ and $R_0 \lhd R$ are normal subgroups and moreover that the embedding of $C$ in $L \times R$ is precisely the graph of an isomorphism $L/L_0 \cong R/R_0$.  Indeed, given $a \in L$ one sends it to some $b \in R$ such that $(a,b) \in C$.  There is no unique such $b$, but any two are in the same $R_0$ coset, whence this process defines a map $L \to R/R_0$, which is surjective by the definition of $R$ and can be shown to be a group homomorphism.  The kernel of this map is clearly $L_0$, which proves the desired isomorphism.  This result is known as \textbf{Goursat's lemma}.

Now back to our geometric problem, we have a finite subgroup $\Gamma$ of $\Sp(1) \times \Sp(1)$ with projections $L,R$ which are finite subgroups of $\Sp(1)$.  We argued that for the action on the sphere to be free the subgroups $L_0$ and $R_0$ have to be trivial, whence Goursat's lemma says that $L \cong R$ and hence that $\Gamma$ embeds in $L \times L$ as the graph of an automorphism $\alpha$ of $L$.  Of course, taking $\alpha$ to be the identity, we recover precisely the groups discussed in the section on $N=5$.  Similarly, if $\alpha$ is an inner automorphism, hence induced by conjugation, then this gives conjugate subgroups in $\SO(8)$ and hence the quotient is isometric to that when $\alpha$ is the identity, leading us back to the case of $N=5$.  Hence in order to obtain a true $N=4$ quotient we need only consider outer automorphisms\footnote{We must own up to a slight abuse of notation, by which automorphisms which are not inner will be referred to as outer, even though strictly speaking outer automorphisms are equivalence classes of automorphisms modulo inner automorphisms.} of the finite subgroups of $\Sp(1)$.  It is however not guaranteed that outer automorphisms will give rise to inequivalent quotients.  The relevant condition is whether or not the effect of the automorphism can be undone via conjugation in $\SO(8)$ or in $\Spin(8)$ in the spinor representation.  This seems to boil down to whether the automorphism in question can be realised by conjugation in $\SO(4)$.  In the quaternionic picture, an element $u \in \Sp(1)$ is embedded into $\SO(\HH)\cong \SO(4)$ as the linear transformation $q \mapsto u q$, for $q \in \HH$.  The element $[(u_L,u_R)] \in \SO(\HH)$ acts via $q \mapsto u_L q \bar u_R$, for $(u_L,u_R) \in \Sp(1) \times \Sp(1)$, so it acts on linear transformations by conjugation: the endomorphism $(q \mapsto uq)$ in $\SO(\HH)$ transforms into $q \mapsto u_L (u (\bar u_L q u_R))\bar u_R = u_L u \bar u_L q$, whence $u \mapsto u_L u \bar u_L$ gets conjugated in $\Sp(1)$.  The question is thus whether the normaliser $N(\Gamma)$ of $\Gamma$ in $\Sp(1)$ acts on $\Gamma$ in such a way that it surjects on the full automorphism group.  If it does then all automorphisms are realised as conjugation in $\SO(8)$ and hence the quotients are isometric.  For the case of $2T$ the outer automorphisms are in fact induced by conjugating with the normaliser, but this is not the case for $2O$ nor $2I$.

In summary, the possible $N=4$ quotients are of the following form.  Let $\Gamma < \Sp(1)$ be a finite group and let $\iota: \Gamma \to \SO(4)$ denote the restriction to $\Gamma$ of the embedding defined by equation \eqref{eq:sp1inso4}.  If $\alpha \in \Aut(\Gamma)$ is an automorphism, we will let $\iota^\alpha : \Gamma \to \SO(4)$ denote the embedding defined by $\iota^\alpha(u) = \iota(\alpha(u))$.  Putting the two together, we have an embedding $\Gamma \into \SO(8)$ defined by the sequence of embeddings
\begin{equation}
  \label{eq:twist}
  \begin{CD}
      \Gamma @>{\iota \times \iota^\alpha}>> \SO(4) \times \SO(4) @>>> \SO(8),
  \end{CD}
\end{equation}
the second one being the obvious one.  In other words, if $u \in \Gamma$, then its action on $(q_1,q_2) \in \HH^2 \cong \RR^8$ is $u\cdot (q_1,q_2) = (uq_1,\alpha(u)q_2)$.  If $\alpha$ is the identity we will call this the \textbf{diagonal} embedding of $\Gamma$ in $\SO(8)$, otherwise we will say the embedding is \textbf{twisted} (by $\alpha$).  As mentioned above if $\alpha$ is an inner automorphism, then the twisted embedding is conjugate in $\SO(8)$ to the diagonal embedding and hence give rise to isometric quotients.  Hence to classify the possible $N=4$ quotients we will first determine the possible outer automorphisms of $\Gamma$, then we will determine the possible lifts to $\Spin(8)$ of the subgroup of $\SO(8)$ obtained by embedding $\Gamma$ as in equation \eqref{eq:twist} and for each such outer automorphism and each such lift we will calculate the dimension of the space of invariant spinors.

This latter task will be accomplished using character formulae.  To compute $N_\pm = \dim\rsc^{\Gamma}$, the dimensions of the $\Gamma$-invariant subspaces of chiral spinors, we use the projection formula
\begin{equation}
  \label{eq:projection}
  N_\pm = \tfrac1{|\Gamma|} \sum_{u \in \Gamma} \chi_{\rsc}(u)~,
\end{equation}
where $\chi_{\rsc}(u)$ is the trace of the group element $u$ in the representation $\rsc$, respectively.  We can work directly in the Clifford algebra $\Cl(8) \cong \RR(16)$ and by going to an explicit matrix realisation for $\Cl(8)$ where the $\gamma_i$ are real $16\times 16$ matrices, then the group elements $u$ will also be $16\times 16$ matrices.  Then their characters are given simply by
\begin{equation}
  \label{eq:trace}
  \chi_{\rsc}(u) = \Tr u \Pi_\pm~,
\end{equation}
where $\Pi_\pm = \half(1 \pm \gamma_9)$ are the projectors corresponding to the chiral spinor representations.

Since characters are class functions, the calculation is further simplified by summing not over the group but over conjugacy classes:
\begin{equation}
  \label{eq:projcc}
  N_\pm = \tfrac1{|\Gamma|} \sum_{\text{conjugacy classes}~[u]} |[u]|\times\chi_{\rsc}(u)~,
\end{equation}
where $|[u]|$ is the size of the conjugacy class of $u$.

We will perform this analysis type by type, starting with the cyclic subgroups.

\subsection{$\sA_{n-1}$}
\label{sec:an}

For the group of type $\sA_{n-1}$, with $n\geq 2$, which is cyclic of order $n$, all automorphisms are outer, since the group is abelian.  An automorphism is determined uniquely by what it does to a generator, and being an automorphism it has to send it to another generator.  Hence the group of automorphisms is the group $\ZZ^\times_n$ of multiplicative units in $\ZZ_n$; that is, those integers in the range $\{1,\ldots,n-1\}$ which are coprime to $n$.  The order of $\ZZ^\times_n$ is the value $\phi(n)$ of Euler's totient function.

Let $r\in \ZZ_n^\times$ act on $\ZZ_n$ by sending a generator $t$ to $t^r$.  Then the generator $t=e^{i2\pi/n}$ of $\ZZ_n$ lifts to $\Spin(8) \subset \Cl(8)$ as
\begin{equation}
  \hat t = \sigma \exp\frac{2\pi}{n} (\I + r \I')~,
\end{equation}
in terms of the elements of $\Cl(8)$ introduced in equation \eqref{eq:IJK}, and where $\sigma$ is a sign satisfying $\sigma^n = 1$, whence if $n$ is odd then $\sigma =1$.  This is precisely what one expects from the isomorphisms in \eqref{eq:spinstructs}.  Here $\ZZ_{n}$ is abelian, whence the spin structures in any (smooth, at least) quotient are in one-to-one correspondence with
\begin{equation}
  \Hom(\ZZ_n,\ZZ_2) \cong
  \begin{cases}
    \ZZ_2, & \text{if $n$ is even;}\\
    \{1\}, & \text{if $n$ is odd.}
  \end{cases}
\end{equation}
Hence there are two spin structures if $n$ is even (labelled by $\sigma$ above) and only one if $n$ is odd.

Using an explicit matrix realisation for $\Cl(8)$ we can compute the characters of the spinor representations:
\begin{align}
  \chi_\rs(t^p) &= \Tr_\rs(\hat t^p) = 2\sigma^p \left( 2 + \cos\frac{2p(r-1)\pi}{n} + \cos\frac{2p(r+1)\pi}{n} \right)\\
  \intertext{and}
  \chi_\rc(t^p) &= \Tr_\rc(\hat t^p) = 4 \sigma^p \left( \cos\frac{2p\pi}{n} + \cos\frac{2p r \pi}{n} \right)~.
\end{align}

Let us first compute $N_-$.  The projection formula \eqref{eq:projection} says that
\begin{align*}
  N_- &= \frac{1}{n} \sum_{p=0}^{n-1} 4 \sigma^p \left(\cos\frac{2p\pi}{n} + \cos\frac{2pr\pi}{n}\right)\\
  &= \frac{2}{n} \sum_{p=0}^{n-1} \left( (\sigma e^{\frac{i2\pi}{n}})^p + (\sigma e^{\frac{-i2\pi}{n}})^p + (\sigma e^{\frac{-i2r\pi}{n}})^p + (\sigma e^{\frac{-i2r\pi}{n}})^p \right)~.
\end{align*}
Now, each of the sums is geometric:
\begin{equation}
  \sum_{p=0}^{n-1} z^p =
  \begin{cases}
    \frac{z^n-1}{z-1} & z\neq 1\\
    n & z = 1~,
  \end{cases}
\end{equation}
and moreover the expression for $z$ is such that $z^n=1$.  Hence the sum gives either $0$ if $z \neq 1$ or else $n$.   We need to identify the cases with $z=1$.  If $\sigma=1$, this happens when $e^{\frac{i2\pi x}{n}}=1$, for $x= \pm 1, \pm r$, which requires $x = 0 \pmod{n}$.  Clearly this can never happen, whence if $\sigma=1$, we find that $N_-=0$.  If $\sigma=-1$, which requires $n$ even and hence $r$ odd, $z=1$ happens when $e^{\frac{i2\pi x}{n}}=-1$, which requires $2x/n$ to be an odd integer.  Clearly if $n>2$ this cannot happen, since $r$ is coprime to $n$, but if $n=2$, then this happens in all four cases $x=\pm1, \pm r$.  In this case we have $N_-=8$, which corresponds to the other spin structure on $\RP^7$.

To compute $N_+$ we again use the projection formula \eqref{eq:projection} to obtain
\begin{align*}
  N_+ &= \frac{1}{n} \sum_{p=0}^{n-1} 2 \sigma^p \left( 2 + \cos\frac{2p(r-1)\pi}{n} + \cos\frac{2p(r+1)\pi}{n} \right)\\
  &= \frac{1}{n} \sum_{p=0}^{n-1} \left( 4 \sigma^p + (\sigma e^{\frac{i2(r-1)\pi}{n}})^p + (\sigma e^{\frac{-i2(r-1)\pi}{n}})^p + (\sigma e^{\frac{i2(r+1)\pi}{n}})^p + (\sigma e^{\frac{-i2(r+1)\pi}{n}})^p  \right) 
\end{align*}
The sums are again geometric, whence the need to identify the cases when the summand is $1$.  Looking first at the first summand, we see that if $\sigma=1$ this contributes $4$ to $N_+$, whereas if $\sigma = -1$, which implies that $n$ is even, the first sum cancels.  We may summarise this by saying the first sum gives a contribution of $2(1+\sigma)$.  The rest of the sums depend on what $r$ is.  For $r$ generic, the exponential sums vanish, e.g.,
\begin{equation}
  \sum_{p=0}^{n-1} (\sigma e^{\frac{i2(r\pm 1)\pi}{n}})^p  = \frac{\sigma^{n} e^{i2(r\pm 1)\pi} - 1}{e^{\frac{i2(r\pm1)\pi}{n}}-1} = 0~,
\end{equation}
since $\sigma^n=1$.  Hence the generic such quotient has $N_+=4$ for the positive spin structure and $N_+=0$ for the negative spin structure.  For some values of $r$, however, the exponentials sum might contribute.

Let us first of all consider the case of $\sigma = 1$.  We have to distinguish the cases $n>2$ and $n=2$.  If $n>2$ then if $r = \pm 1 \pmod{n}$, we have that precisely two of the exponential sums contribute an extra $2$ to $N_+$, bringing it up to $N_+=6$.  Notice that for $n=3,4,6$ there are no other possible values of $r$ than these.  If $n=2$, then $r$ is odd, since it is coprime to $n$, and hence $r\pm1$ is even, in which case all four exponential sums contribute equally yielding $N_+=8$.  Of course, $r=1$ is the diagonal embedding and $r=-1$ corresponds to a twisted embedding which is conjugate (by $j$, say) in $\Sp(1)$ to the diagonal one, whence the quotients are isometric.

Finally let us consider the case of $\sigma = -1$, which requires $n$ to be even.  Then the exponential sums will contribute provided that $r\pm 1 = \frac{n}{2} \pmod{n}$.  Since $r$ has to be coprime to $n$, it has to be odd, which then forces $n$ to be a multiple of $4$.  In this case precisely two of the sums contribute and we find $N_+ = 2$.

In summary, to obtain an $N_+=4$ quotient we need to take $\sigma = 1$ and $r \neq \pm 1 \pmod{n}$.

\subsection{$\sD_{n+2}$}
\label{sec:dn}

This is the binary dihedral group $2D_{2n}$, for $n\geq 2$, defined abstractly in terms of generators and relations as
\begin{equation}
  \label{eq:2Dn}
  2D_{2n} = \left<s,t \middle| s^2 = t^n = (st)^2\right>~,
\end{equation}
or explicitly in terms of quaternions by $s = j$ and $t = e^{i\pi/n}$, with central element $s^2 = t^n = (st)^2 = -1$.  The group has order $4n$.  There are $n+3$ conjugacy classes which are listed along with their sizes and the orders of their elements in Table \ref{tab:cc2D}, which also displays the characters of the spinorial representations.  The notation $\ell(p,2n)$ in that table means the least common multiple of $p$ and $2n$.

Automorphisms are uniquely determined by their action on generators, provided that their images still satisfy the relations.  In addition, automorphisms must preserve the order of the elements and must also map conjugacy classes to conjugacy classes.  By definition, inner automorphisms preserve the conjugacy classes, so if an automorphism does not then it must be outer.  These considerations nail down the outer automorphisms once the conjugacy classes are enumerated, as we have done.  First we can consider the automorphisms which fix $s$ and transform $t \mapsto t^r$ where $(r,2n)=1$.  Notice that $t^r$ still obeys $(t^r)^n = -1$, since $r$ is odd, and that $(st^r)^2=-1$, whence $s$ and $st^r$ are again generators.  Since the conjugacy class $\{t,t^{-1}\}$ gets mapped to $\{t^r,t^{-r}\}$, this automorphism is outer for $r\neq 1,2n-1$.  Also we can consider the automorphism which fixes $t$ and sends $s \mapsto s t$.  We could consider a more general automorphism sending $s \mapsto st^q$ but up to inner automorphisms (which fix $t$) only $q=1$ need be considered.  This is an automorphism because $(st)^2 = (st^2)^2 = -1$, so that the relations are satisfied.  It is clearly outer, since $s$ and $st$ belong to different conjugacy classes.  These two automorphisms commute and clearly they give rise to a group of outer automorphisms isomorphic to $\ZZ_{2n}^\times \times \ZZ_2$.  We observe that the element $j e^{i\pi/2n} \in \Sp(1)$ conjugates $st$ back to $s$ at the price of sending $t$ to $t^{-1}$, which is an equivalent generator.  Hence we can always ignore the $\ZZ_2$ factor from the point of view of obtaining quotients with $N=4$.  In other words, from now on we will consider the outer automorphisms which leave $s$ fixed and send $t \mapsto t^r$ for $r \in \ZZ_{2n}^\times\setminus \{\pm 1\}$.

Depending on the parity of $n$, we have either 2 or 4 possible spinorial lifts of $2D_{2n}$.   This follows from the isomorphisms in \eqref{eq:spinstructs} once we understand the abelianisation of $2D_{2n}$.  It is not hard to see that the commutator subgroup is the cyclic subgroup generated by $t^2$, which has order $n$.  This means that the abelianisation has order $4$ and hence can either be cyclic or isomorphic to Klein's \emph{Viergruppe}, a.k.a. $\ZZ_2 \times \ZZ_2$.  To determine which, we argue as follows.  If $n$ is even, then $-1$ is in the commutator subgroup, so modulo the commutator subgroup the four elements $1$, $s$, $t$ and $st$ surject onto the abelianisation: since both $s$ and $t$ have order $2$ modulo the commutator subgroup, this four-element group is isomorphic to $\ZZ_2 \times \ZZ_2$.  On the other hand, if $n$ is odd, then $-1$ is not in the commutator subgroup, but modulo the commutator subgroup $t = -1$. Hence the four elements which surject to the abelianisation are $\pm1$ and $\pm s$, which is a cyclic group of order $4$ with generator $s$.  Finally, using \eqref{eq:spinstructs} and the isomorphisms
\begin{equation}
  \Hom(\ZZ_4,\ZZ_2) \cong \ZZ_2 \qquad\text{and}\qquad \Hom(\ZZ_2\times\ZZ_2,\ZZ_2) \cong \ZZ_2 \times \ZZ_2~,
\end{equation}
we expect to see two signs labelling the lifts of $2D_{2n}$ to $\Spin(8)$ when $n$ is even and only one sign when $n$ is odd.

Indeed, let us first consider the lift to $\Cl(4)$.  In the notation of equation \eqref{eq:IJK}, the following generate a group of $\Cl(4)$ isomorphic to $2D_{2n}$:
\begin{equation}
    \hat s = \sigma \exp\frac{\pi}{2}\J \qquad\text{and}\qquad \hat t = \tau \exp\frac{\pi}{n}\I~,
\end{equation}
where $\sigma$ and $\tau$ are signs, but with $\tau^n=1$.  The lifts to $\Spin(8)$ corresponding to the embedding twisted by $r \in \ZZ_{2n}^\times$ is therefore given by
\begin{equation}
    \hat s = \sigma \exp \frac{\pi}{2} (\J + \J')  \qquad\text{and}\qquad \hat t = \tau \exp \frac{\pi}{n} (\I + r\I')~,
\end{equation}
with $\sigma,\tau$ signs again with $\tau^n=1$.  It is now straightforward to compute the spinorial characters by using an explicit realisation of $\Cl(8)$.  The results are given in Table~\ref{tab:cc2D}, in which $p=1,...,n-1$.

\begin{table}[h!]
  \caption{Conjugacy classes and spinorial characters of $2D_{2n}$}
  \centering
  \begin{tabular}{>{$}c<{$}|>{$}c<{$}|>{$}c<{$}|>{$}c<{$}|>{$}c<{$}}
    \text{Class} & \text{Size} & \text{Order} & \chi_{\rs} & \chi_{\rc} \\\hline
    1 & 1 & 1 & 8 & 8 \\
    -1 & 1 & 2 & 8 & -8\\
    s & n & 4 & 4\sigma & 0 \\
    st & n & 4 & 4\sigma\tau & 0 \\
    t^p & 2 & \ell(2n,p)/p & 2\tau^p(2 + \cos\frac{p(r-1)\pi}{n} + \cos\frac{p(r+1)\pi}{n}) & 4 \tau^p (\cos\frac{p\pi}{n} + \cos\frac{p r \pi}{n})
  \end{tabular}
  \label{tab:cc2D}
\end{table}

Using the projection formula \eqref{eq:projcc}, we determine $N_\pm$.  First of all, as a check of our calculations, we should obtain that $N_-=0$.  This is very similar in spirit to the computation of $N_-$ in the cyclic case.  Indeed, one has
\begin{equation*}
  \begin{split}
    N_- &= \frac{1}{4n} \left( 8 - 8 + 0 + 0 + 2 \sum_{p=1}^{n-1} 4 \tau^p \left( \cos \frac{p\pi}{n} + \cos \frac{pr\pi}{n} \right) \right)\\
    &= \frac{1}{n} \sum_{p=1}^{n-1} \left( (\tau e^{i\pi/n})^p + (\tau e^{-i\pi/n})^p + (\tau e^{ir\pi/n})^p + (\tau e^{-ir\pi/n})^p \right)\\
    &= \frac{1}{n} \left( \frac{\tau^n + e^{i\pi/n}\tau}{1-\tau e^{i\pi/n}} + \frac{\tau^n + e^{-i\pi/n}\tau}{1-\tau e^{-i\pi/n}} + \frac{-e^{ir\pi} \tau^n + e^{ir\pi/n}\tau}{1-\tau e^{ir\pi/n}} + \frac{-e^{-ir\pi}\tau^n + e^{-ir\pi/n}\tau}{1-\tau e^{-ir\pi/n}} \right).
\end{split}
\end{equation*}
Using that $r$ is odd and that $\tau^n=1$, we can rewrite this expression as
\begin{equation*}
  \begin{split}
   N_- &= \frac{1}{n} \left( \frac{1 + e^{i\pi/n}\tau}{1-\tau e^{i\pi/n}} + \frac{1 + e^{-i\pi/n}\tau}{1-\tau e^{-i\pi/n}} + \frac{1 + e^{ir\pi/n}\tau}{1-\tau e^{ir\pi/n}} + \frac{1 + e^{-ir\pi/n}\tau}{1-\tau e^{-ir\pi/n}}\right)\\
    &= \frac{1}{n} \left( \frac{1 + e^{i\pi/n}\tau}{1-\tau e^{i\pi/n}} + \frac{e^{i\pi/n}\tau + 1}{\tau e^{i\pi/n}-1} + \frac{1 + e^{ir\pi/n}\tau}{1-\tau e^{ir\pi/n}} + \frac{e^{ir\pi/n}\tau + 1 }{\tau e^{ir\pi/n}-1}\right) = 0,
  \end{split}
\end{equation*}
where in the last line we multiplied the second term top and bottom by $\tau e^{i\pi/n}$ and the fourth term by $\tau e^{ir\pi/n}$ and used that $\tau^2=1$.

Moving on to the computation of $N_+$, the projection formula says that
\begin{equation*}
  \begin{split}
    N_+ &= \frac{1}{4n} \left(8 + 8 + 4n\sigma + 4n\sigma\tau + \sum_{p=1}^{n-1} 4\tau^p \left( 2 + \cos \frac{p(r-1)\pi}{n} + \cos \frac{p(r+1)\pi}{n}\right)\right)\\
    &= \sigma(1+\tau) + \frac{1}{n} \sum_{p=0}^{n-1} \tau^p \left( 2 + \cos \frac{p(r-1)\pi}{n} + \cos \frac{p(r+1)\pi}{n}\right)\\
    &= (1 +\sigma)(1 + \tau)  + \frac{1}{2n} \sum_{p=0}^{n-1} \left( (\tau e^{\frac{i(r-1)\pi}{n}})^p + (\tau e^{-\frac{i(r-1)\pi}{n}})^p + (\tau e^{\frac{i(r+1)\pi}{n}})^p + (\tau e^{-\frac{i(r+1)\pi}{n}})^p \right).
  \end{split}
\end{equation*}
For generic values of $r$, the exponential sums vanish:
\begin{equation}
  \sum_{p=0}^{n-1} (\tau e^{\pm\frac{i(r\pm1)\pi}{n}})^p = \frac{\tau^n e^{\pm i (r\pm 1)\pi} - 1}{\tau e^{\pm\frac{i(r\pm1)\pi}{n}}-1} = 0~,
\end{equation}
since $\tau^n=1$ and $r$ is odd since it is coprime to $2n$.  In these cases we have $N_+ = (1+\sigma)(1+\tau)$, whence $N_+ = 4$ if $\sigma=\tau=1$ and $N_+=0$ otherwise.  For some values of $r$, for which $\tau e^{\frac{i(r\pm1)\pi}{n}} = 1$, then two of the corresponding exponential sums will contribute a total of $1$ to $N_+$.  

Let us first consider $\tau=1$.  Then we have extra contributions whenever
\begin{equation}
  \exp\left(\frac{i(r\pm 1)\pi}{n}\right) = 1,
\end{equation}
whence if $r = \pm 1 \pmod{2n}$, then we have $N_+=5$ for $\sigma=1$ and $N_+=1$ for $\sigma=-1$.  It should be pointed out that for $n=2$ and $n=3$ there are no $r\in\ZZ_{2n}^\times$ which do not obey $r = \pm 1 \pmod{2n}$, whence there are no $N_+=4$ quotients for these values of $n$.  Finally let us consider the case of $\tau=-1$, which forces $n$ to be even.  Then we get extra contributions whenever
\begin{equation}
  \exp\left(\frac{i(r\pm 1)\pi}{n}\right) = -1~,
\end{equation}
which implies that $r\pm 1$ is an \emph{odd} multiple of $n$, hence $r= n\pm 1$.  In this case we have that $N_+=1$ regardless the value of $\sigma$.

In summary, we get $N_+=4$ for $\sigma=\tau=1$ and $r \neq \pm 1\pmod{2n}$, which requires $n\geq 4$.  We get $N_+=5$ for $\sigma=\tau=1$ and $r = \pm 1 \pmod{2n}$.

\subsection{$\sE_6$}
\label{sec:e6}

The group of type $\sE_6$ is the binary tetrahedral group $2T$ of order $24$.  Abstractly it is isomorphic to $\SL(2,\FF_3)$.  It has a centre of order 2 and quotienting by it gives the tetrahedral group $T \cong \PSL(2,\FF_3) \cong A_4$.  This last isomorphism can be understood because $\PSL(2,\FF_3)$ acts faithfully on the projective space $\PP_1(\FF_3)$ of lines through the origin in $\FF_3$, consisting of four points, so it embeds in $S_4$ as a subgroup of index 2 and there is only one such subgroup.  The full automorphism group of $2T$ is the symmetric group $S_4$ and a representative outer automorphism can be obtained by conjugation in $\GL(2,\FF_3)$ with an element not in $\SL(2,\FF_3)$.  To relate this to the above description, notice that the group $\GL(2,\FF_3)$ acts faithfully on $\PP_1(\FF_3)$ whence it embeds in $S_4$, but given its order it has to be all of $S_4$.

We can understand this outer automorphism also in terms of the usual presentation
\begin{equation}
  2T = \left<s,t\middle| s^3 = t^3 = (st)^2\right>~,
\end{equation}
where
\begin{equation}
  s = \half (1 + i + j + k) \qquad\text{and}\qquad  t = \half (1 + i + j - k).
\end{equation}
The group $2T$ has seven conjugacy classes, tabulated below in Table~\ref{tab:cc2T}, along with its size and the order of any (and hence all) of its elements.

\begin{table}[h!]
  \centering
  \caption{Conjugacy classes of $2T$}
  \begin{tabular}{*{8}{>{$}c<{$}|}}
    \text{Class} & 1 & -1 & s & t & t^2 & s^2 & st \\
    \text{Size} & 1 & 1 & 4 & 4 & 4 & 4 & 6 \\
    \text{Order} & 1 & 2 & 6 & 6 & 3 & 3 & 4 \\
  \end{tabular}
  \label{tab:cc2T}
\end{table}

We can see that the only nontrivial outer automorphism is one which sends $s$ to the conjugacy class of $t$ and viceversa.  In fact, since $st$ and $ts$ are conjugate, it is possible to represent the nontrivial outer automorphisms by the automorphism which exchanges the two generators.  This automorphism consists of conjugation by $\frac1{\sqrt{2}} (i+j) \in \Sp(1)$.  This means that the twisted embedding of $2T$ in $\SO(8)$ is conjugate in $\SO(8)$ to the diagonal embedding and hence the quotients are isometric.  In particular the twisted embedding gives $N=5$, which can also be checked explicitly; although we will refrain from doing so here.  Let us nevertheless remark that there is a unique spin structure in the quotient, because the commutator subgroup of $2T$ has index $3$, whence the abelianisation is isomorphic to $\ZZ_3$, but $\Hom(\ZZ_3,\ZZ_2) = \{1\}$.

\subsection{$\sE_7$}
\label{sec:e7}

The group of type $\sE_7$ is the binary octahedral group $2O$ of order $48$.  The automorphism group is isomorphic to $O \times \ZZ_2$, where $O$ is the octahedral group, which is the quotient of $2O$ by its centre, whence it is clear that the factor $O$ corresponds to the inner automorphisms.  The group admits the following presentation
\begin{equation}
  2O = \left<s,t\middle| s^3 = t^4 = (st)^2\right>~,
\end{equation}
and we can make this explicit in terms of quaternions by choosing
\begin{equation}
  s = \half (1 + i + j + k) \qquad\text{and}\qquad t = \frac1{\sqrt{2}} (1 + i)~.
\end{equation}
There are 8 conjugacy classes tabulated in Table~\ref{tab:cc2O} below, which also contains the computation of the characters, described in more detail below.

\begin{table}[h!]
  \centering
  \caption{Conjugacy classes and spinorial characters of $2O$}
  \begin{tabular}{>{$}c<{$}*{4}{|>{$}c<{$}}}
    \text{Class} & \text{Size} & \text{Order} & \chi_{\rs} & \chi_{\rc} \\\hline
    1 & 1 & 1 & 8 & 8\\
   -1 & 1 & 2 & 8 & -8\\
    s & 8 & 6 & 5 & 4\\
    t & 6 & 8 & 2 \sigma(2+\tau) & 2 \sqrt{2}\sigma(1+ \tau)\\
    s^2 & 8 & 3 & 5 & -4\\
    t^2 & 6 & 4 & 4 & 0 \\
    t^3 & 6 & 8 & 2 \sigma(2+\tau) & -2 \sqrt{2}\sigma(1+ \tau)\\
    st & 12 & 4 & 4\sigma & 0 \\
  \end{tabular}
  \label{tab:cc2O}
\end{table}

Since automorphisms preserve orders, we see that any automorphism must send $s$ to its conjugacy class, hence modulo inner automorphisms we can let $s$ remain fixed.  The other generator $t$ has to be sent to the other conjugacy class of the same size and order to that of $t$.  We see from the table that it must be to the conjugacy class of $t^3$.  It cannot be sent to $t^3$, however, because $st^3$ has order 8 and not 4.  A little playing around suggest defining the automorphism by sending $t \mapsto t^5 = -t$ and fixing $s$. One checks that the relations are satisfied.  It is not difficult to show that this automorphism cannot be obtained by conjugation in $\Sp(1)$.

We now investigate the lift of the generators to $\Spin(4) \subset \Cl(4)$ as a preliminary stage to lifting them via the twisted embedding to $\Spin(8)$.  First of all, from equation \eqref{eq:spinstructs} we can already see that there are two inequivalent spin structures on any quotient.  This is because the commutator subgroup of $2O$ has index $2$, whence the abelianisation is isomorphic to $\ZZ_2$ and $\Hom(\ZZ_2,\ZZ_2) \cong \ZZ_2$.  So we expect the lift to depend on a sign.

Indeed, in the notation of \eqref{eq:IJK}, we find the following lifts
\begin{equation}
    \hat s = \exp \frac{\pi}{3} 3^{-\frac12} (\I + \J + \K ) \qquad\text{and}\qquad \hat t_\tau = \sigma \exp\frac{\pi}{4} (2\tau-1) \I~,
\end{equation}
where $\sigma$ and $\tau$ are signs: $\sigma$ labels the two inequivalent lifts, whereas $\tau$ distinguishes between the lift $\hat t_+$ of $t$ and that $\hat t_-$ of $-t$.  The diagonal and the twisted embeddings give $2O$ subgroups of $\SO(8)$ acting freely on the unit sphere.  For each of these two embeddings we have two possible lifts to $\Spin(8)$, corresponding to the two spin structures of the quotient.  The lifts are explicitly given by
\begin{equation}
    \hat s = \exp \frac{\pi}{3} 3^{-\frac12}(\I + \J + \K + \I' + \J' + \K')  \qquad\text{and}\qquad \hat t = \sigma \exp \frac{\pi}{4}(\I + (2\tau-1) \I')~,
\end{equation}
where $\tau=1$ is the diagonal embedding and $\tau = -1$ the twisted embedding and $\sigma$ labels the two spin structures.

It is now a simple matter to use the trace formulae \eqref{eq:trace} in an explicit representation of the Clifford algebra in order to compute the characters of $2O$ in the spinor representations for arbitrary $\sigma$ and $\tau$.  The result of this calculation is found in Table \ref{tab:cc2O}.  The dimension of the invariant subspaces is easily computed from the tabulated data using the projection formula \eqref{eq:projcc} and one finds that whereas $N_-=0$ for both embeddings, we have
\begin{equation}
  N_+ = \half (5 + \sigma (4 + \tau)).
\end{equation}
Therefore for the diagonal embedding ($\tau = 1$), we have $N_+ = 5$ for the positive spin structure ($\sigma =1$) and $N_+ = 0$ for the negative spin structure, providing a check of our calculations. More interestingly, for the twisted embedding ($\tau = -1$) we find $N_+=4$ for the positive spin structure and $N_+=1$ for the negative one.

\subsection{$\sE_8$}
\label{sec:e8}

Finally, the group of type $\sE_8$ is the binary icosahedral group $2I$ of order 120, which is isomorphic to $\SL(2,\FF_5)$.  The automorphism group is the symmetric group $S_5$.  This is similar in spirit to the case of $\sE_6$, in that the representative outer automorphism can be obtained via conjugation in $\GL(2,\FF_5)$.  A  representative for the outer automorphism using this method is given in \cite[§2.2.5]{SunilThesis}.  We will find another representative here by exploiting the  structure of the group.

The usual presentation is
\begin{equation}
  2I = \left<s,t\middle| s^3 = t^5 = (st)^2\right>~,
\end{equation}
and we may take $s$ and $t$ to be the quaternions
\begin{equation}
  s = \half (1 + i + j + k) \qquad\text{and}\qquad t = \half (\varphi + \varphi^{-1} i + j)~,
\end{equation}
where $\varphi = \half (1 + \sqrt{5})$ and $\varphi^{-1} = \half (-1 + \sqrt{5})$ are the Golden Ratio and its reciprocal, respectively.  The group has 9 conjugacy classes which are tabulated in Table \ref{tab:cc2I}  along with their size and order and the spinorial characters in both the diagonal and twisted embeddings.

\begin{table}[h!]
  \caption{Conjugacy classes and spinorial characters of $2I$}
  \centering
  \begin{tabular}{>{$}c<{$}*{6}{|>{$}c<{$}}}
     &&& \multicolumn{2}{c|}{Diagonal} & \multicolumn{2}{c}{Twisted}\\
    \text{Class} & \text{Size} & \text{Order} & \chi_{\rs} & \chi_{\rc} & \chi_{\rs} & \chi_{\rc} \\\hline
    1 & 1 & 1 & 8 & 8 & 8 & \phantom{-}8 \\
   -1 & 1 & 2 & 8 & -8 & 8 & -8\\
    t & 12 & 10 & 5+\varphi & 4\varphi & 3 &  \phantom{-}2 \\
    t^2 & 12 & 5 & 5-\varphi^{-1} &4 \varphi^{-1} & 3 & -2 \\
    t^3 & 12 & 10 & 5-\varphi^{-1} & -4\varphi^{-1} & 3  &  \phantom{-}2 \\
    t^4 & 12 & 5 & 5 + \varphi &-4\varphi & 3 & -2 \\
    s & 20 & 6 & 5 & 4 & 5 &  \phantom{-}4 \\
    s^4 & 20 & 3 & 5 & -4 & 5 & -4 \\
    st & 30 & 4 & 4 & 0 & 4 &  \phantom{-}0 \\
  \end{tabular}
  \label{tab:cc2I}
\end{table}

We see from the description of the conjugacy classes that the generator $s$ belongs to the unique class of size 20 and order 6, whence any automorphism must send it to another element in its class, which means that we can undo the effect of that automorphism using an inner automorphism.  Hence modulo inner automorphisms, we are free to leave $s$ alone.  The generator $t$ belongs to one of two conjugacy classes of size 12  and order 10.  An outer automorphism must send it to the class of $t^3$, but alas we cannot send it to $t^3$ itself because $st^3$ is not conjugate to $st$.  A little experimentation suggests that we send $t \mapsto t' := t^6 s t^4 s t^3$. One checks that this is an outer automorphism and moreover one checks that it cannot obtained by conjugation in $\Sp(1)$.

Lifting the elements $s,t,t'$ to $\Cl(4)$ one finds a unique lift, whence both the diagonal and twisted quotients have a unique spin structure.  Of course, this is as expected because $2I$ is perfect, so that the commutator subgroup is all of $2I$ and hence the abelianisation is trivial.  A little calculation shows that the pairs of elements $\{\hat s, \hat t\}$ and $\{\hat s, \hat t'\}$ in $\Spin(4) \subset \Cl(4)$, defined below in the notation of equation \eqref{eq:IJK}, generate separately a group isomorphic to $2I$ and surjecting to $2I \subset \SO(4)$:
\begin{equation}
  \begin{aligned}[m]
    \hat s &= \exp \frac{\pi}{3} 3^{-\frac12}(\I + \J + \K ) \\
    \hat t &= \exp \frac{\pi}{5} \varphi^{-\frac12} 5^{-\frac14} ( \I + \varphi \J)\\
    \hat t' &= \exp \frac{3\pi}{5} \varphi^{-\frac12}5^{-\frac14} ( \K - \varphi \I )~.
  \end{aligned}
\end{equation}

The diagonal and the twisted embeddings give $2I$ subgroups of $\SO(8)$ acting freely on the unit sphere.  For each of these two embeddings we have a unique lift to  $\Spin(8)$.  The lift for the diagonal embedding is
\begin{equation}
  \begin{aligned}
    \hat s &= \exp \frac{\pi}{3} 3^{-\frac12}(\I + \J + \K + \I' + \J' + \K')\\
    \hat t &= \exp \frac{\pi}{5} \varphi^{-\frac12} 5^{-\frac14} ( \I + \varphi \J + \I' + \varphi \J')~,
  \end{aligned}
\end{equation}
whereas for the twisted embedding we have the same $\hat s$ but now
\begin{equation}
     \hat t = \exp \frac{\pi}{5} \varphi^{-\frac12} 5^{-\frac14} ( \I + \varphi \J - 3 \varphi \I' + 3 \K')~.
\end{equation}

It is now a simple matter to use the trace formulae \eqref{eq:trace} in an explicit representation of the Clifford algebra in order to compute the characters of $2I$ in the spinor representations for both the diagonal (as a check) and the twisted embeddings.  The result of this calculation is found in Table \ref{tab:cc2I}.  The dimension of the invariant subspaces is easily computed from the tabulated data and one finds that, whereas $N_-=0$ for both embeddings, one checks that $N_+=5$ for the diagonal embedding and $N_+=4$ for the twisted embedding, giving a new $N=4$ quotient and concluding their classification.

Notice that the Lie algebra of isometries of an $N=4$ quotient is $\fso(4)$, which is not of big enough dimension to act (locally) transitively on a 7-dimensional manifold.  Therefore $N=4$ quotients are not homogeneous.  This agrees with the expectation \cite{FMPHom,JMF-HC-Lecs} that homogeneity is only implied by preserving more than one half of the supersymmetry; that is, for $N>4$.  For the case of riemannian manifolds admitting Killing spinors, this has recently been proved in \cite{ACKilling}.

\section{Summary}
\label{sec:summary}

Table~\ref{tab:quotients} summarises the classification of smooth quotients of $S^7$ of type $(N,0)$ with $N>3$.  We list groups by their ADE labels, but we insist that they are not abstract groups, but the specific subgroups of $\SO(8)$ described above.  In each case there is a unique lift to $\Spin(8)$ with the indicated value of $N$.  The notation $(\mathsf{\Gamma},\alpha)$ means a pair consisting of a finite subgroup of $\Sp(1)$ of type $\mathsf{\Gamma}$ and an outer automorphism $\alpha$.  If $\alpha$ is the identity, we omit it and write only $\mathsf{\Gamma}$.  In the cyclic or binary dihedral $N{=}4$ quotients, the automorphism denoted $r \in \ZZ^\times_m$ (for the appropriate $m$) is the one sending the generator $t$ to $t^r$.  Finally, the automorphism $\nu$ for $\sE_{7,8}$  is any automorphism representing the unique nontrivial outer automorphism.

\begin{table}[h!]
  \caption{Smooth quotients of $S^7$ of type $(N,0)$ with $N>3$}
  \setlength{\extrarowheight}{3pt}
  \centering
  \begin{tabular}{>{$}c<{$}|>{$}l<{$}}
    N & \multicolumn{1}{c}{Groups}\\\hline
    8 & \sA_1\\
    6 & \sA_{n\geq 2}\\
    5 & \sD_{n\geq 4}, \sE_6, \sE_7, \sE_8\\
    4 & (\sA_{n\geq 4, \neq5}, r \in \ZZ^\times_{n+1}\setminus \{\pm1\}), (\sD_{n\geq6}, r \in \ZZ^\times_{2(n-2)}\setminus \{\pm1\}), (\sE_7, \nu), (\sE_8, \nu)
  \end{tabular}
  \label{tab:quotients}
\end{table}
  
\section*{Acknowledgments}

PdM was supported by a Seggie-Brown Postdoctoral Fellowship of the School of Mathematics of the University of Edinburgh.  The work of JMF was supported by a World Premier International Research Center Initiative (WPI Initiative), MEXT, Japan.  JMF takes pleasure in thanking Hitoshi Murayama for the invitation to visit IPMU and the Leverhulme Trust for the award of a Research Fellowship freeing him from his duties at the University of Edinburgh.  EME would like to thank the Simons Workshop in Mathematics and Physics 2009 for their support and hospitality during the completion of this work.

\bibliographystyle{utphys}
\bibliography{Sugra,Geometry,Algebra}

\end{document}